# Simulation studies of the self-assembly of cone-shaped particles

Ting Chen[1], Zhenli Zhang[1] and Sharon C. Glotzer[1,2,*]

[1]Department of Chemical Engineering and [2]Department of Materials Science & Engineering

University of Michigan, Ann Arbor, Michigan 48109-2136

[*]email: sglotzer@umich.edu.

Submitted 08/27/06

We systematically investigate the self-assembly of anisotropic cone-shaped particles decorated by ring-like attractive "patches". We demonstrate that the self-assembled clusters, which arise due to the conical particle's anisotropic shape combined with directional attractive interactions, are precise for certain cluster sizes, resulting in a precise packing sequence of clusters of increasing sizes with decreasing cone angles. We thoroughly explore the dependence of cluster packing on cone angle and cooling rate, and categorize the resulting structures as "stable" and "metastable" clusters. We also discuss the implication of our simulation results in the context of the Israelachvili packing rule for surfactants, and a recent geometrical packing analysis on hard cones in the limit of large numbers of cones.

## I.  INTRODUCTION

Nature has mastered the self-assembly of building block subunits into complicated structures with extraordinary precision and accuracy. Examples range from the assembly of complementary strands of DNA into a double helix, to proteins self-assembling into spherical virus capsids with icosahedral symmetry, to the formation of filaments, fibers and membranes in cells. In these examples, nature exploits the use of highly specific and directional interactions to achieve precisely ordered structures. Mimicking such interactions in synthetic nanostructures by exploiting anisotropy both in building block shape and inter-building block interactions may lead to the fabrication of precise structures for use in a range of applications.

This work is inspired by recent experimental interest in the assembly of cone-like particles and molecules. Examples include the assembly of cone-shaped amphiphilic dendro-calixarenes into precise micelles [1], middle-functionalized cone-like peptide-amphiphiles forming nanofibers [2], and self-assembled superstructures by rod-like metal-polymer amphiphiles [3]. Additionally, some capsomers (morphological subunits composed of groups of proteins) in virus protein capsids [4, 5] possess short, truncated cone-like shapes and, to a first approximation, can be modeled as cone-shaped particles. On larger scales, novel synthetic methods have been developed to make conical colloids. Sheu *et al.* fabricated uniform nonspherical particles from homo-IPNs (interpenetrating polymer networks), including ice cream cone-shaped particles via seeded emulsion polymerization of styrene-divinylbenzene mixtures in crosslinked monodisperse polystyrene seed latexes [6, 7]. Most recently, Douliez [8] demonstrated the synthesis of micron-sized hollow cones of controlled angles via self-assembly in bola-amphiphile/hexadiamine salt solutions. Although progress has been made in the preparation of conical building blocks, the principles underlying their self-assembly into ordered structures are not well understood. As such, it is desirable that a general, predictive theoretical or simulation approach be formulated to study this problem and provide design principles for the assembly of conical building blocks.

A second motivation for this study comes from theoretical considerations of the conical particle packing problem. In an attempt to develop guidance for the assembly of amphiphilic surfactants, Israelachvili [9] described the conditions under which spherical micelles or cylindrical micelles will form, based on a "critical packing parameter" (CPP) or "shape factor". The CPP is defined as $V/a_0 l_c$ where $V$ is the volume of the amphiphilic particle or molecule, $a_0$ is the surface area of the head group and $l_c$ is the critical length. According to the Israelachvili packing rule, when CPP $\leq 1/3$, as in ideal cones, amphiphiles assemble into spherical micelles; when $1/3 <$ CPP $\leq 1/2$, as in truncated cones, cylindrical micelles form. Though the Israelachvili packing rule is successful in rationalizing the self-assembled structures formed in surfactant systems, the rule provides primarily qualitative insight and has limited application. These limitations were recently addressed by Tsonchev *et al.*, who developed a geometric packing analysis for the packing problem of amphiphilic nanoparticles treated as hard cones [10]. With the assumption that all particles pack locally in a hexagonal arrangement, Tsonchev *et al.* predicted that spherical micelles are always preferred in the self-assembly of hard conical particles because of the higher packing fraction in spherical clusters than in cylindrical clusters, in contrast to the commonly held belief that truncated cones will form cylindrical micelles[9]. However, the key assumption of local hexagonal packing in Tsonchev *et al.*'s theory is only valid in the limit of very large $N$, where $N$ is the number of cones in a single cluster. For clusters comprised of smaller numbers of particles, the finite curvature of the assembled cluster will prevent extended hexagonal order [11], altering the cluster shape in an unknown way. For example, as we will show, twelve cone-shaped particles with an angle of 62° form a perfect icosahedral cluster where every cone base has only five nearest neighbors (pentagonal packing) instead of six (hexagonal packing). Furthermore, both the Israelachvili packing rule and the geometric packing analysis of Tsonchev *et al.* focus on the explanation of overall cluster shape and not the local packings of the particles within the clusters.

Recently, Rapaport [12] reported a molecular dynamics (MD) study on the self-assembly of polyhedral shells to investigate the dynamics underlying protein shell formation in spherical viruses. He demonstrated an extension of his method to study rigid tapered cylindrical particles (i.e., cones), and showed several large spherical micelles formed by many tapered cylindrical particles in MD simulations. The present simulation work focuses on the local packing for small to intermediate cluster sizes and also examines the validity of the above predictions for the packing of conical building blocks at large cluster size limit.

In previous work [13], we showed that the self-assembled structures from cone-shaped particles belong to a more general packing sequence that includes polyhedral structures formed by evaporation-driven assembly of colloidal microspheres [14] and several virus capsid structures. In this article, we extend our investigation and tackle key issues unaddressed in our previous work, and discuss the influence of cone angle on cluster packings, the cluster size distribution and yield, and the influence of cooling rate on the assembled structures. Additionally, we discuss the implications of our simulation results as they pertain to the Israelachvili packing rule for surfactants, and the geometrical packing analysis on hard cones.

## II. MODEL AND SIMULATION METHOD

Rigid cone-shaped particles are constructed by fusing together a linear array of overlapping spherical subunits, or beads, with decreasing sizes. Varying the number of beads and/or the inter-bead distance controls the length of a cone, and cone angle is controlled by varying the size gradient along the axis of the array.

In this work, we consider six-bead cone-shaped particles in which the distance between neighboring beads within a cone is $0.5\sigma$, where $\sigma$ is the diameter of the smallest bead. As shown in FIG. 1, the cone angle $\theta$ can be calculated from the equation $\sin(\frac{\theta}{2}) = \frac{r_6 - r_1}{2.5}$, where $r_6$ and $r_1$ are the radii of the largest and smallest bead in the particle, respectively. The constant 2.5 is the distance between the centers of the two end beads. Square-well interactions of well depth $\varepsilon$ exist between beads occupying the same positions (indicated by the same colors in FIG. 1) on the particles except for the two end beads. The end beads, and unlike beads with different colors, interact through a hard-core excluded volume interaction only. The interaction range (width of square well) $\lambda$ is fixed at $0.4\sigma$. The particle-based simulation approach we employ here has the advantage that it makes no *a priori* assumptions regarding local packing.

Monte Carlo (MC) simulation in the canonical ensemble using the standard Metropolis algorithm is used to simulate the systems. The simulation box is cubic and periodic boundary conditions are implemented in all three Cartesian coordinates. At each MC step, either a translational or a rotational move is attempted for each particle [15] and the trial move is accepted by comparing the Boltzmann factor, which is calculated from the potential energy difference between the configurations before and after the trial move, to a computer generated random number. The ratio of attempted translational moves to rotational moves is 2 : 8. Initially, all cone-shaped particles (roughly 500 ~ 1000 particles at a volume fraction of approximately 0.1 ~ 0.3) are randomly distributed throughout the simulation box at a high temperature. The maximum translational displacement is chosen to be $0.2\sigma$ and the maximum rotational displacement is chosen to be 0.2 rad.  We choose a cooling rate that is affordably slow yet fast enough to produce stable structures reliably and efficiently. The influence of cooling rate is also investigated. The system is slowly cooled to a low, target reduced temperature (0.3 ~ 0.5) to allow sufficient time for the particles in the system to assemble into ordered structures. The cone angle is systematically varied (with small increments 0.5 ~ 1°) in multiple simulations to study the dependence of cluster packings on cone angle. In addition, multiple independent runs at certain cone angles are performed with different initial configurations and along different cooling paths to investigate the path dependence of the structures. The cluster size distribution information is collected by using a previously developed analysis code [16].

# III.  RESULTS AND DISCUSSION

This section is organized as follows. In section A, we describe the structural characteristics for each precise cluster we obtained at certain "magic number" values $N$. In section B, we describe how we determine the cone angle range for each precise structure and discuss the stability of clusters based on the cone angle range. Section C discusses cluster size distribution. Section D discusses the influence of cooling rate on the cone angle range and cluster size distribution. Finally, we discuss our simulation results for large cluster size in the context of previous theoretical studies in section E.

## A. Cluster structures at small to moderate $N$

As we systematically vary the cone angle, we find upon cooling that a series of unique clusters are obtained with a specific number of cones, $N$. Each size $N$ cluster has a range of cone angles over which that $N$-particle cluster can be assembled. FIG. 2 lists the visual images of structures at $N$ = 4 - 17, 20, 27, 32 and 42, along with the corresponding cone angle ranges. The collection of such information results in a diagram for the assembly of cone-shaped particles, as shown in FIG. 3. The cooling rate used is $\Delta T = -0.01$ per 0.5 million MCS (Monte Carlo steps). The influence of cooling rate on the cone angle ranges is discussed later in section E. The packings observed in different simulations are robust and independent of cone angle.

The directional attraction between cone-shaped particles is the driving force for assembly. Particles attempt to maximize their number of nearest neighbors to achieve the lowest energy state possible at the target temperature. At the same time, the conical excluded volume forces the particles to assemble into a curved, closed structure. The specific lateral attractions betweens cones expedite their assembly into tightly bound clusters. Once in the cluster, the outermost beads of the cones are effectively confined on the surface of a convex shell formed by the inner beads of the cones. FIG. 2 shows clusters with small sizes ($N \leq 17$) have distinct polyhedral convex shapes.

Platonic solids, packings of spheres, low energy Lennard-Jones clusters, and equilibrium configurations of point charges on a spherical surface all relate to different types of polyhedral cluster structures [17]. While our precise packings do share some common features with some of these polyhedral structures, they also demonstrate some interesting, unique characteristics. For example, if we take the head bead (the largest bead) in a cone as a vertex and construct a convex hull formed by all vertices (head beads) within a cluster, clusters with size $N = 4$ -10 and 12 shown in FIG. 2 are the eight convex deltahedra with 4, 6, 8, 10, 12, 14, 16 and 20 faces, $i.e.$, tetrahedron ($N = 4$), triangular dipyramid ($N = 5$), octahedron ($N = 6$), pentagonal dipyramid ($N = 7$), snub disphenoid ($N = 8$), triaugmented triangular prism ($N = 9$), gyroelongated square dipyramid ($N =10$) and icosahedron ($N = 12$) [18]. The definition of a deltahedron is a polyhedron whose faces consist of congruent equilateral triangles that are not in the same plane. The same packings were also observed in experiments on colloidal microspheres that formed clusters upon droplet evaporation; in that paper the relationship of convex deltahedra to other applications was discussed [14]. The investigation of evaporation-driven assembly of colloidal spheres was discussed in our previous work [13]. At the smallest possible cone angle for the $N = 8$ clusters, we observed a twisted-square structure instead of the snub-disphenoid. This twisted-square structure was observed in evaporation experiments by Cho *et al.* [19, 20] while the snub-disphenoid structure was observed in experiments by Manoharan *et al.* [14] and Yi *et al.* [21]

The $N = 11$ cluster is, however, special. It is not a deltahedron because no 18-face deltahedron exists. Since one of its vertices is shared by six triangles, there should be a hexagonal packing plane if all six triangles are equilateral. This is clearly not the case for the $N = 11$ cluster as shown in FIG. 2. Through the vertex that has six neighbors, there is a vertical reflection plane. Similar to the $N = 11$ cluster, the $N = 13$ cluster also has a vertical reflection plane through the cluster center. Both clusters lack high symmetry as compared to the $N = 12$ cluster with icosahedral symmetry.

The $N = 14$ cluster has a biplanar structure with $D_6$ symmetry ($D_n$ symmetry refers to dihedral rotational symmetry, or rotational symmetry with mirror symmetry) with some additional subtlety. Each plane of the $N = 14$ cluster contains seven hexagonally arranged cone-shaped particles, but with opposite rotation direction and two planes in the staggered conformation. In one plane, seven cone-shaped particles rotate clockwise with respect to the center particle into the $R$ (*rectus*) configuration while in the other plane they rotate counterclockwise into the $S$ (*sinister*) configuration. This same packing was also observed in the evaporation-driven assembled structures of 14 colloidal microspheres but without the inherent chirality observed here [14].

The $N = 20$ cluster has a short cylinder-like shape. It can be best described as three adjacent layers stacked together with $D_{6h}$ symmetry ($D_{nh}$ symmetry refers to dihedral rotational symmetry with reflective symmetry in a horizontal mirror). Each of the top and bottom layers contains seven particles arranged hexagonally in an eclipsed conformation and the middle layer is a six-particle ring in a staggered conformation with both top and bottom layers.

The $N = 27$ cluster has $D_{5h}$ symmetry. The 12 pentamers are distributed into a 1:5:5:1 arrangement and have five particles in the horizontal reflection plane. The two halves are in an eclipsed conformation. This cluster structure may correspond to a non-icosahedral spherical virus structure, such as the middle component of the pea enation mosaic virus, the top component of the tobacco streak virus, and the Tulare apple mosaic virus, which consist of roughly 150 subunits or 27 capsomers (12 pentamers and 15 hexamers), and violate the Caspar-Klug (CK) quasi-equivalence theory, as suggested by Cusack [22]. In contrast, the $N = 32$ cluster has icosahedral symmetry with 12 pentamers and 20 hexamers. Viruses with a triangulation number $T = 3$ have 32 capsomers in their capsid and exhibit the same icosahedral symmetry as shown here. $C_{60}$ fullerenes also have the same symmetry that consists of 12 pentagonal faces and 20 hexagonal faces.

The $N = 42$ cluster, which is the largest precisely packed structure that we obtain within a reasonable computation time, does not have the expected icosahedral symmetry [23]. Interestingly, the $D_{5h}$ symmetry of this non-icosahedral structure is also predicted in the proposed optimal packings of the maximum volume of a convex hull for a set of points on a sphere[24]– a closely related mathematical problem – and in the packings of point charges on a sphere, known as the Thomson problem in mathematics [23].

It is worth noting that all the packings at the discussed values of $N$ ($N = 4 - 17$, 20, 27, 32 and 42) in the presented sequence are unique and robust, therefore we refer to them as "magic number" clusters, in analogy with Lennard-Jones clusters [25]. Clusters of size $N$ other than these values exhibit multiple polymorphs and hence are not unique. These clusters are also usually characterized by low symmetries. As an example, FIG. 4 shows three different 22-particle structures formed via self-assembly of cone-shaped particles with a cone angle of 43.4°.

## B. Cone angle ranges

### *1. Method used to identify the cone angle ranges.*

We estimate the range of angles, defined as the difference of upper limit and lower limit of cone angles that will form clusters of a given size, by analyzing the cluster size distribution at each investigated cone angle. The upper limit of the cone angle capable of producing a cluster of a given size can be directly estimated from the cluster size distribution. Since we vary the cone angle in small increments (about 0.5 - 1°), it is easy to locate the largest angle at which a given cluster size disappears. We then take the last cone angle that still yields this cluster size as the upper bound on its cone angle range.

The determination of the lower limit of cone angle capable of producing a given cluster size is less straightforward. At these angles the system is usually a mixture of complete clusters and incomplete clusters that will eventually form slightly larger clusters. We then use visual inspection combined with the information deduced from the cluster size distribution to determine the lower limit of cone angle that can self-assemble into clusters of a given size. Again, we take the smallest possible cone angle that still can form a given size of cluster as the lower boundary of its cone angle range.

The error of determining cone angle boundaries for a cluster of given size is limited only by the angle steps we used ($0.5 \sim 1°$) in the simulations. Statistical runs for typical clusters $N = 11$, 12 and 13 demonstrate that the fluctuation of cone angle boundaries determined by this method varies within 1 degree, as shown in FIG. 3.

The simulation-estimated range of cone angles for each structure, which can be considered as a kind of "tolerance", shows whether it is possible for cone-shaped particles with a given angle to assemble into clusters of a given size.

### 2. "Stable" and "metastable" clusters.

The range of angles over which a given "magic number" cluster structure is obtained is an indication of the stability of that structure at the current set of simulation parameters, and thus we can categorize the $N$-particle clusters as "stable" and "metastable" clusters according to their cone angle ranges. We note that this classification, while quantitative, is subjective and not based on evaluation of the cluster free energy.

FIG. 3 shows that the angle ranges can differ significantly and change in a discontinuous way with respect to $N$. The overall tendency is for the angle range to decrease as $N$ increases. However, some clusters have unusually small or large angle ranges as compared to adjacent ones. For example, the $N = 4$ and 6 clusters have the widest angle ranges of 51.2° and 28.4°, respectively, while both $N = 5$ and 7 clusters have angle ranges of only about 10°. Similarly, the cone angles ranges for the $N = 9$ and 12 clusters are significantly larger than the adjacent $N = 11$ and 13 clusters.

The potential energy part of the free energy decreases monotonically as $N$ increases, as the cone particles in larger clusters have more neighboring particles of attractive interactions on average. However, entropy is not a monotonic function of $N$. The effect of entropy on cluster stability may be understood from symmetry arguments. An icosahedron possesses 6 five-fold, 10 three-fold and 15 two-fold rotational symmetries through its vertices, faces and edges. Such high symmetry represents high configurational entropy and low free energy, assuming relatively unvaried or insignificant potential energy. As such, a 12-particle cluster with an icosahedral symmetry is particularly stable, and can be found assembled from conical particles with a wide range of angles. In contrast, the $N = 11$ cluster contains one particle fewer than the most stable icosahedron cluster and is of substantially lower symmetry, and thus lower entropy. As such, the $N = 11$ cluster has an unusually small cone angle range indicating low stability.

In summary, we categorize the clusters into "stable" (like $N = 4, 6, 9, 12$) and "metastable" clusters (like $N = 5, 7, 11$ and 13), characterized by wide and narrow cone angle ranges, respectively. The configurational entropy is likely to be the cause of the discontinuous change in stability with respect to the cluster size $N$.

## C. Cluster size distribution and yield

The cluster size distribution gives important information about the yield, or the possibility of obtaining clusters of a certain size as well as the relative stability among clusters with comparable sizes at a given cone angle. FIG. 5 shows typical cluster size distributions for $N = 11$, 12, 13, 14, and 32 at their optimum angles (that give the highest yield of a specific cluster size at a cooling rate of $\Delta T = -0.01$ per 0.5 million MCS) of 67.4°, 62°, 55.7° and 35.52°, respectively.

Since the $N = 12$ cluster has extremely stable icosahedral symmetry, the size distribution at the optimum cone angle, 62°, is monodisperse, as shown in FIG. 5(b). In contrast, the yields of the $N = 11$ and 13 clusters are low as expected, typically under 10%, where the $N = 11$ cluster is peculiar, as shown in FIG. 5(a), because its angle range and yield is particularly small, indicating its low stability compared to other cluster sizes.

As we decrease the cone angle $\theta$, more cone-shaped particles can self-assemble into clusters with larger sizes. The difficulty of obtaining complete clusters in the simulation usually increases as $\theta$ decreases, i.e., as the cluster size increases. As such, "stable" clusters with small sizes, such as $N = 4$ and $N = 6$ clusters, display monodisperse distributions within a range of angles (102.5° to 116.4° for the former and 81° to 92° for the latter). In contrast, "metastable" clusters like $N = 5$ and $N = 7$ clusters have a relatively broad distribution and low yield even at their optimum angles. For other stable clusters, such as the $N = 8$ and $N = 9$ cluster, they also have narrow distribution and relatively high yield (over 70%).

It is worth noting that both 13- and 14-particle clusters share the same optimum cone angle as 55.7° at the employed cooling rate. But the fact that the latter distribution has a much higher peak than the former in FIG. 5(c) shows the $N = 14$ cluster is more stable that the $N = 13$ cluster. For the other two clusters with comparable sizes, $N = 11$ and $N = 12$ clusters, the $N = 12$ icosahedral cluster is always favored over the $N = 11$ cluster under all conditions. FIG. 5(d) shows a relatively high yield of 32-particle icosahedral clusters for cone-shaped particles with an angle of 35.52°, though the size distribution is relatively broad.

To conclude this part, "stable" clusters like $N = 4$, 6 and 12 have nearly monodisperse distribution and high yield in contrast to relatively broader cluster size distribution and low yield for "metastable" clusters like $N = 11$ and 13, consistent with their corresponding wide and narrow angle ranges in the previous discussion.

In the evaporation experiments on colloidal microspheres [14, 21], the cluster size is set by the number of particles trapped in a droplet. There is no direct control over the structures as they form and separation processes are required to obtain clusters of a particular size. In contrast, with a properly chosen cone angle it is possible to control how many particles can assemble into a single cluster, and with pre-determined, desired precise packing. We have also predicted that precisely packed structures with a narrow size distribution can be assembled from cone-shaped particle with certain cone angles, which is of technical significance because it may substantially reduce the cost of, if not eliminate the need for, the separation procedures required in assembly of isotropic particle building blocks. This demonstrates the possibility of using patterned cone-shaped particles as a programmable building block for self-assembly.

## D. Influence of cooling rate

*1. Influence of cooling rate on size distribution profile and cluster yield.*

In order to investigate the influence of cooling rates, we compare the size distributions of cone-shaped particles at certain cone angles for three cooling rates. Cooling rate 0 decreases the temperature by 0.02 per 0.5 million MCS and cooling rate 1 decreases the temperature by 0.01 per 0.5 million MCS. Cooling rate 2 decreases the temperature by 0.002 per 0.5 million MCS and cooling rate 3 decreases the temperature by 0.001 per 1 million MCS.

Within certain angle ranges of "stable" clusters, slower cooling rates show little or no influence on the already nearly monodisperse size distribution, such as $102.5°$ to $116.4°$ for the $N = 4$ cluster, $81°$ to $92°$ for the $N = 6$ cluster, and $57.4°$ to $62.7°$ for the $N = 12$ cluster.

For other cone angles, a slower cooling rate shifts the size distribution to the larger $N$ side and increases the yield of larger clusters, as shown in FIG. 6 for $\theta = 45.3°$, $55.7°$ and $67.4°$. One interesting question is: will the size distribution at all angles eventually evolve into a single sharp peak with slow enough cooling rate? The simulations using the slowest cooling rate (cooling rate 3) for two cone systems at $55.7°$ and $67.4°$ still show the coexistence of at least two clusters with comparable sizes, though the slower cooling rate does appear to result in fewer peaks in the size distribution profile.

Again, we emphasize that all the packings at each "magic number" $N$ are robust and are independent of the cone angles at which they are observed, e.g., 14-particle clusters at 50.3° and 55.7° are identical, although the yield of the cluster for each angle may change with different cooling rates.

### 2. Influence of cooling rate on cone angle ranges.

The cone angle ranges for clusters of size $N = 6, 9, 12, 13, 14,$ and 20 may be calculated from simulations using different cooling rates. The results are listed in TABLE I.

For "stable" magic number clusters, we observe that the cooling rate has little or no influence on the width of their cone angle ranges. At each cone angle investigated, slow cooling shifts the size distribution to the right side (larger $N$) and produces a higher yield for larger clusters. Since slow cooling shifts both the lower boundary and upper boundary by $0.5 \sim 3$ degrees in the same direction, the net effect is that the cone angle range remains roughly the same, with an overall shift towards larger $N$.

For "metastable" magic number clusters, the influence of cooling rate is, however, complicated. We find that slow cooling rate favors those clusters that are thermodynamically more stable, such as the $N = 4, 6, 12$ clusters, and suppresses, or even diminishes the appearance of those "metastable" clusters, such as the $N = 11$ and 13 clusters. For example, simulations using cooling rate 2 significantly decrease the yields as well as the cone angle ranges of the $N = 11$ and 13 clusters. A faster cooling rate, however, encourages the existence of "metastable" structures. The suppressing effect of slow cooling rate suggests that these clusters are actually stable intermediates, or metastable.

## E. Cluster structures at larger $N$ and implication for previous theoretical studies

In contrast to uniquely packed polyhedral clusters at small to moderate $N$, hard cone-shaped particles form spherical clusters without well-defined local packings at larger $N$ (or at small cone angles). This finding is consistent with previous theoretical prediction [10] and simulation results [12], as shown in FIG. 7, where two larger spherical clusters are obtained with 206 and 475 cone-shaped particles, respectively.

Some possible explanations for this observation are as follows. For hard cones, a spherical cluster has higher packing fraction, or lower potential energy, and thus is more stable or robust than a cylindrical cluster [10]. Additionally, a spherical cluster has higher symmetry and thus larger configurational entropy, which may also contribute to the persistently observed spherical clusters in our cone-shaped particle simulations. As for the local packings, according to Euler's theorem [26], 12 disclinations, i.e., 12 five-fold vertices, are required to close a hexagonal network, and different ways of distributing the 12 disclinations dictate the different local packings inside the spherical cluster, resulting in the protrusions on the surface of the spherical clusters in FIG. 7.

The Israelachvili packing rule is often used to rationalize the structures assembled from surfactants and amphiphilic particles. It is worth noting that no cylindrical clusters are observed in our simulation, in contrast to the prediction of the Israelachvili rule for truncated cones. Our observation, however, may be limited by the rigid cone systems we investigated, while the Israelachvili packing rule was initially conceived for soft amphiphiles. The fact that tubular structures have been frequently observed in experiments on surfactants or amphiphiles [3] implies the limitation of our hard cone model. A soft cone model may explain these experimentally observed cylindrical micelles.

Iacovella *et al*. [27] applied the Israelachvili packing factor [9] to a mono-tethered nanosphere system. They calculated the shape factor from the effective length of the tail (or tether), the effective volume of the tail, and the effective area of the head group (or nanoparticle). For a fixed building block concentration and head diameter, they found transitions from spherical micelles to hexagonally packed cylinders to lamellar bilayers on increasing tether length, which is consistent with the theoretical predictions of Israelachvili. This indicates that the Israelachvili packing rule is more suitable for describing the behavior of amphiphiles of soft conical particles.

A similar situation was also observed by Park *et al*. [3], where rod-like metal-polymer amphiphilic particles were considered as truncated cones with soft, aggregating tails and were qualitatively analyzed using the Israelachvili rule to explain the appearance of self-assembled tubular and sheet structures. Modeling the peptide amphiphile as a lower, rigid part and an upper, flexible part and using molecular dynamics in 2D space,

Tsonchev *et al*. [28] found that long-range dipole interactions in the flexible part and hydrogen bonding in the rigid parts together stabilize peptide amphiphile self-assembly into cylindrical micelles as found in experiments [2]. As such, a more general conical model containing the influence of "softness" of building blocks is promising to encompass more experimental results; our model may be easily extended to meet this challenge.

## IV. CONCLUSIONS

In summary, we investigated the self-assembly of anisotropic rigid cone-shaped particles with directional attractions and demonstrated that structures with pre-designed geometric accuracy could be obtained by the self-assembly of such building blocks. We presented a diagram that can be referenced to predict the self-assembled structure for a given cone angle, providing important guidance for the design and assembly of conical building blocks. Specifically, we found that when $N$, the number of cone-shaped particles in a cluster, is small, the finite packings obtained from self-assembly process produce a series of distinct cluster structures that resemble the packings found in evaporation-driven assembly of colloidal spheres. When $N$ is large, sphere-like clusters, instead of cylindrical clusters as predicted by the Israelachvili rule for truncated cones, are always found, which is consistent with previous predictions by the geometric packing analysis and MD simulations on hard cones

Moreover, we find that certain clusters have narrow cluster size distributions indicating desirable stability and reproducibility, which suggests the superiority of conical building blocks over isotropic building blocks for the fabrication of precise structures. We classify our simulated precise structures into "stable" and "metastable" clusters by their wide and narrow cone angle ranges, and by their monodisperse and non-monodisperse cluster size distribution profiles, and relatively high and low yield.

Additionally, we find that for "stable" clusters, varying cooling rates has little or no influence on their cone angle ranges and size distribution profiles. However, slower cooling rates can change cluster size distribution profiles and cluster yields significantly for "metastable" clusters. Furthermore, a slow cooling rate promotes thermodynamically "stable" clusters and suppresses "metastable" clusters, as expected.

The design principles identified in this work by investigating the dependence of cluster packings and cluster size distribution on cone angles enable programmable and predictable assembly of precise structures with patterned hard cones, and may provide promising routes of fabricating novel designer materials. While the specific angle ranges of stability will undoubtedly change somewhat for cones of different size or different interaction range, the diagram provides an accurate prediction for cones matching the specifications modeled here and general design principles for cones of different specifications. The better understanding achieved in this work on the assembly of anisotropic conical particles may also help explain and control the structure of matter at different length scales for future applications.

Finally, how might such ring-like attractive patches in our model cone be introduced onto cone-shaped particles in experiments? Several experimental approaches appear promising. Sheu *et al*. [6, 7] obtained uniform nonspherical particles such as ice cream cone-like particles from identical polystyrene polymers. It is possible to extend this technique to synthesize cone-like particles from different polymers and convey different properties to the cap and lower part of the ice cream cone-shaped particle, for example, by making the cap to be hydrophilic and the lower part to be hydrophobic. The hydrophobic interaction with solvent may thus drive the assembly of particles into structures such as those studied here. Alternatively, a decorating technique newly developed by Stellacci and coworkers [29] may also be used where ordered phase-separated domains of stabilizing ligands spontaneously occur on the curved surface of nanoparticles.

It is worth mentioning that the model and simulation approach described in this work can be easily extended to study other anisotropic colloidal particles that can be found in recent experiments, such as dumbbell [6, 7, 30], peanut [30, 31], ellipsoid [32], and spindle-like particles [30].

# ACKNOWLEDGEMENTS

This work was supported by the US Department of Energy, Grant No. DE-FG02-02ER46000.

# FIGURE CAPTIONS

**FIG.1.** Illustration of the model cone.

**FIG. 2.** Cluster packings with corresponding cone angle ranges. The cone angle range is shown below each cluster along with the optimum angle for this cluster that gives the highest yield, except that at $N = 8$, two isometric clusters are shown at the two cone angle boundaries. The cooling rate used is $\Delta T = -0.01$ per 0.5 million MCS.

**FIG. 3.** The diagram for the assembly of cone-shaped particles. X axis is the cone angle. Y axis is the cluster size in terms of the number of cone-shaped particles within the cluster. For each cluster in the precise packing sequence, the cone angle ranges represented by thick, color lines at corresponding $N$ give the cone angles over which a $N$-particle cluster is obtained. The cooling rate used is $\Delta T = -0.01$ per 0.5 million MCS. The influence of cooling rate on the cone angle ranges will be discussed later. For $N = 11$, 12 and 13 clusters, the estimated error in the cone angle boundaries are also shown. Note the error bar is omitted for the $N = 11$ cluster because the line that represents the cone angle range of the $N = 11$ cluster is smaller than its error bar.

**FIG. 4.** Examples of polymorphism of non-"magic number" cluster at $N = 22$.

**FIG. 5.** Typical cluster size distributions for cone-shaped particles with an angle of (a) 67.4, (b) 62, (c) 55.7 and (d) 35.52 degrees. Cooling rate: $\Delta T = -0.01$ per 0.5 million MCS.

**FIG. 6.** The influence of cooling rate on cluster size distribution profiles under different cone angles. (a) $\theta = 67.4°$; (b) $\theta = 55.7°$; (c) $\theta = 45.3°$. Cooling rate 1: $\Delta T = -0.01$ per 0.5 million MCS; cooling rate 2: $\Delta T = -0.002$ per 0.5 million MCS; cooling rate 3: $\Delta T = -0.001$ per 1 million MCS.

**FIG. 7.** Images of large, sphere-like clusters self-assembled from cone-shaped particles with small cone angle. (a) A cluster that has 206 particles with a cone angle of 13.8°. (b) A cluster that contains 475 cone-shaped particles with a cone angle of 9.2°. Both clusters have overall round shapes.

**TABLE I.** Influence of cooling rates on the cone angle ranges of clusters with size $N = 6$, 9, 12, 13, 14 and 20. Cooling rate 0: $\Delta T = -0.02$ per 0.5 million MCS; Cooling rate 1: $\Delta T = -0.01$ per 0.5 million MCS; cooling rate 2: $\Delta T = -0.002$ per 0.5 million MCS.

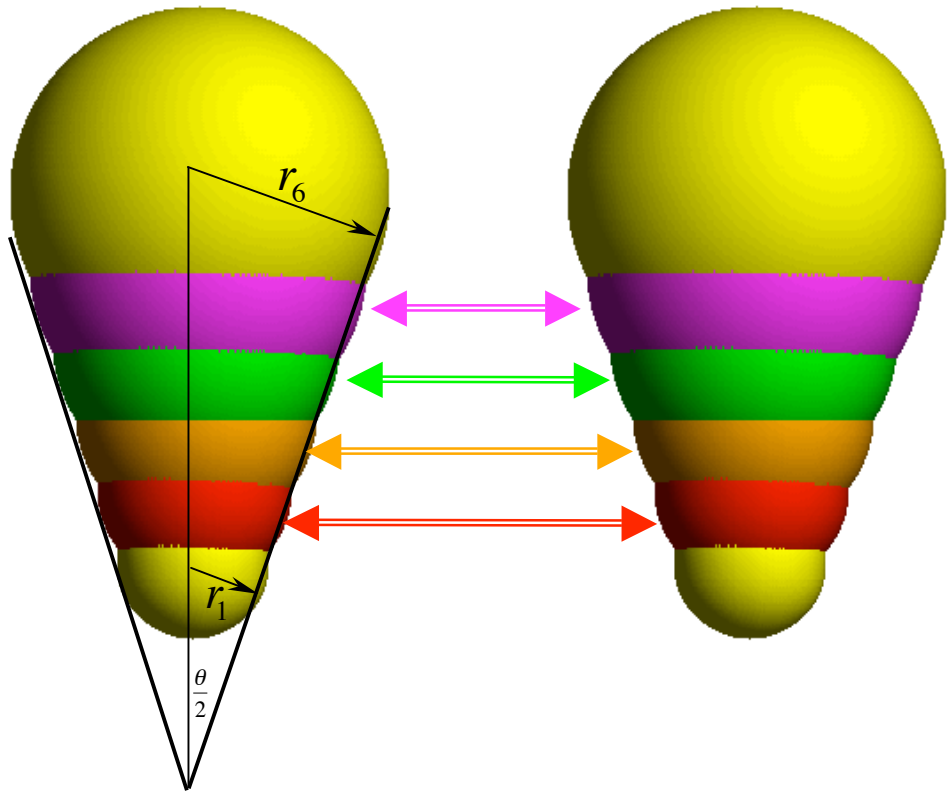

FIG.1.

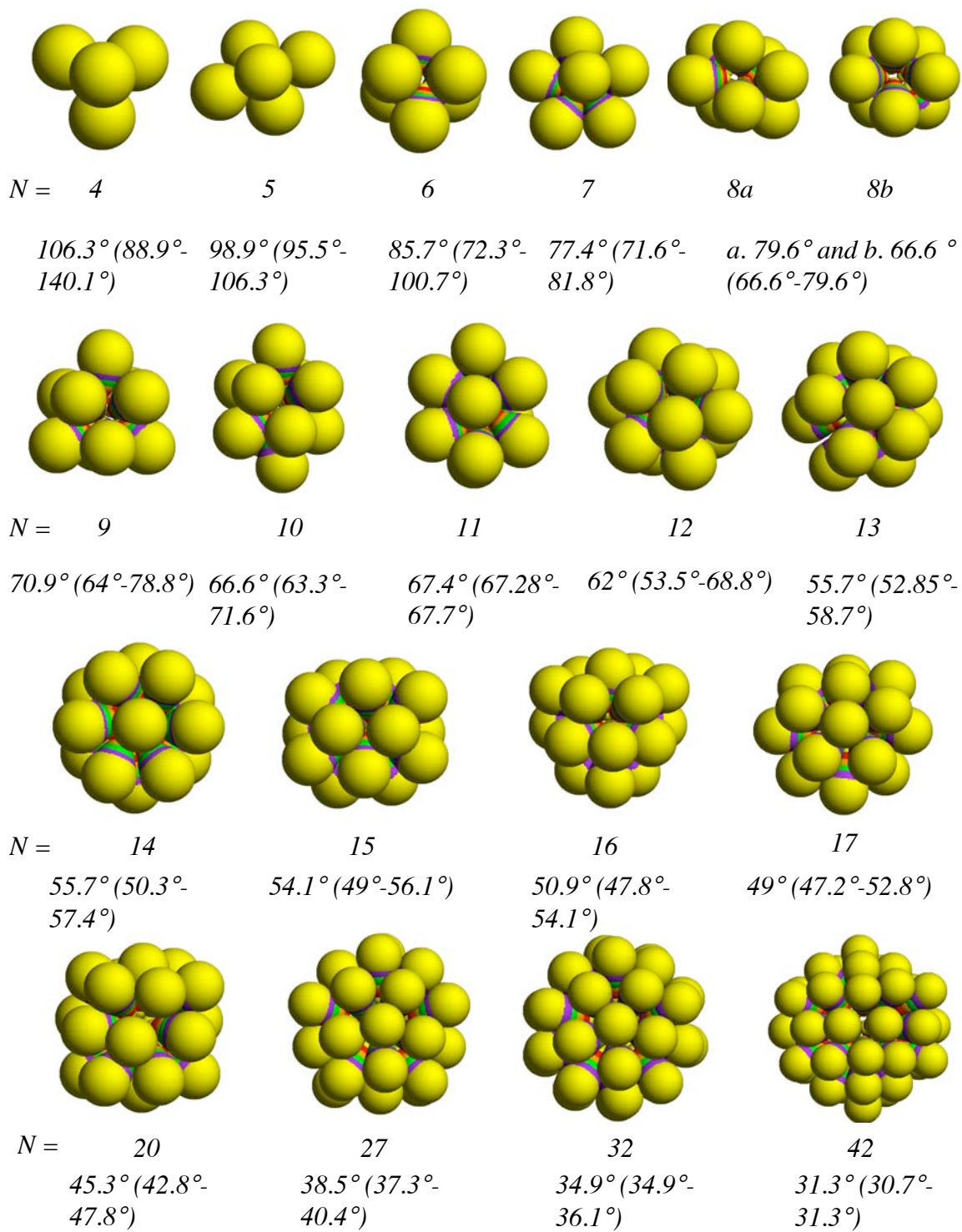

$N =$    *4*         *5*         *6*         *7*         *8a*        *8b*

*106.3° (88.9°-*    *98.9° (95.5°-*    *85.7° (72.3°-*    *77.4° (71.6°-*    *a. 79.6° and b. 66.6 °*
*140.1°)*    *106.3°)*    *100.7°)*    *81.8°)*    *(66.6°-79.6°)*

$N =$    *9*         *10*         *11*         *12*         *13*

*70.9° (64°-78.8°)*    *66.6° (63.3°-*    *67.4° (67.28°-*    *62° (53.5°-68.8°)*    *55.7° (52.85°-*
        *71.6°)*    *67.7°)*        *58.7°)*

$N =$    *14*         *15*         *16*         *17*

*55.7° (50.3°-*    *54.1° (49°-56.1°)*    *50.9° (47.8°-*    *49° (47.2°-52.8°)*
*57.4°)*            *54.1°)*

$N =$    *20*         *27*         *32*         *42*

*45.3° (42.8°-*    *38.5° (37.3°-*    *34.9° (34.9°-*    *31.3° (30.7°-*
*47.8°)*    *40.4°)*    *36.1°)*    *31.3°)*

**FIG.2.**

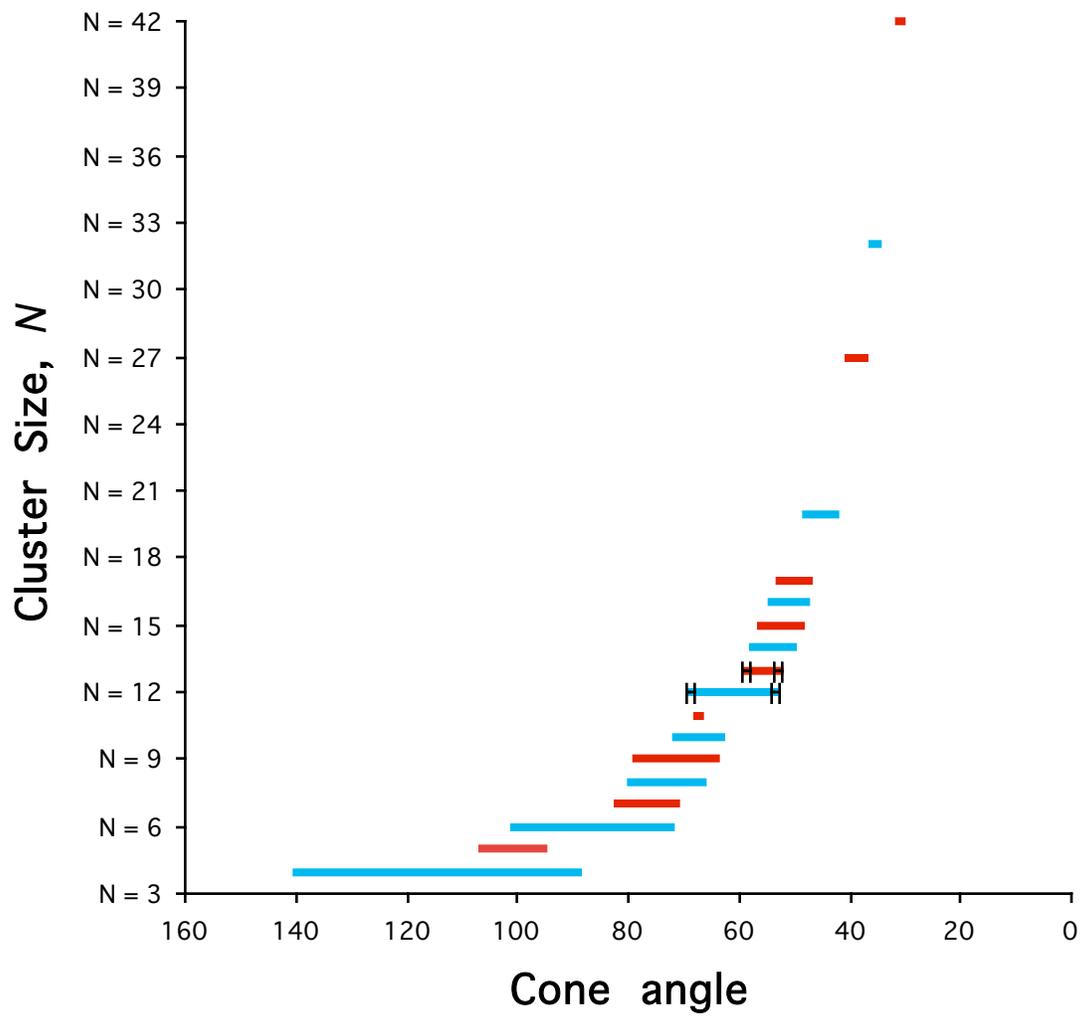

**FIG.3.**

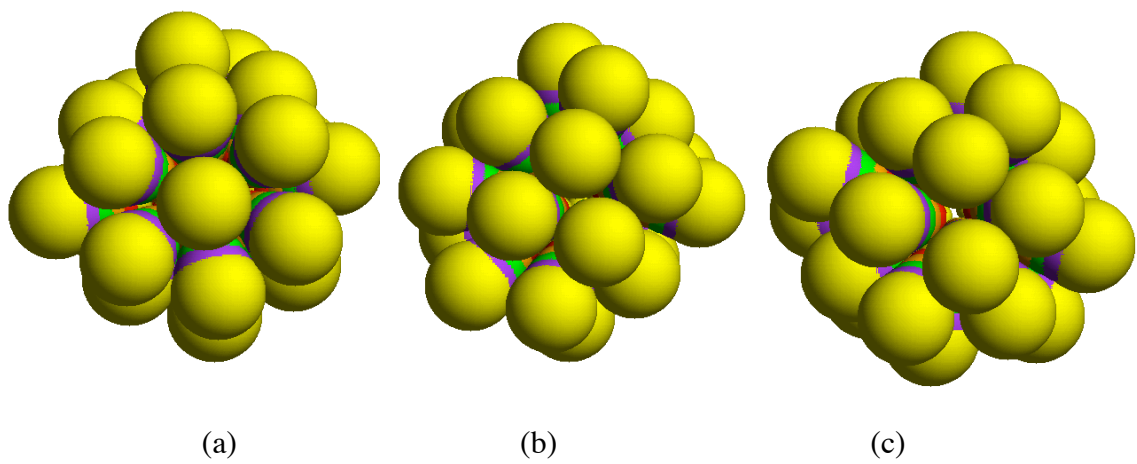

(a)　　　　　(b)　　　　　(c)

**FIG.4.**

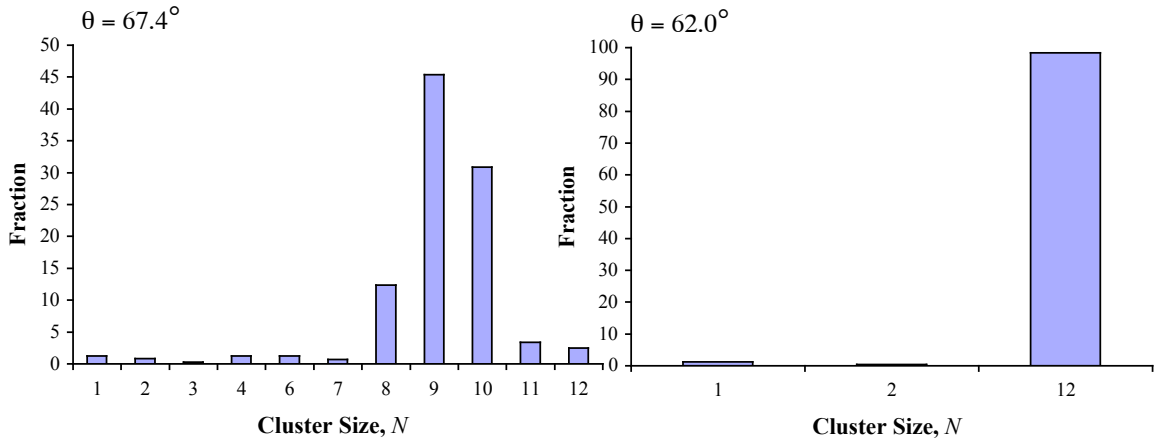

(a)                                           (b)

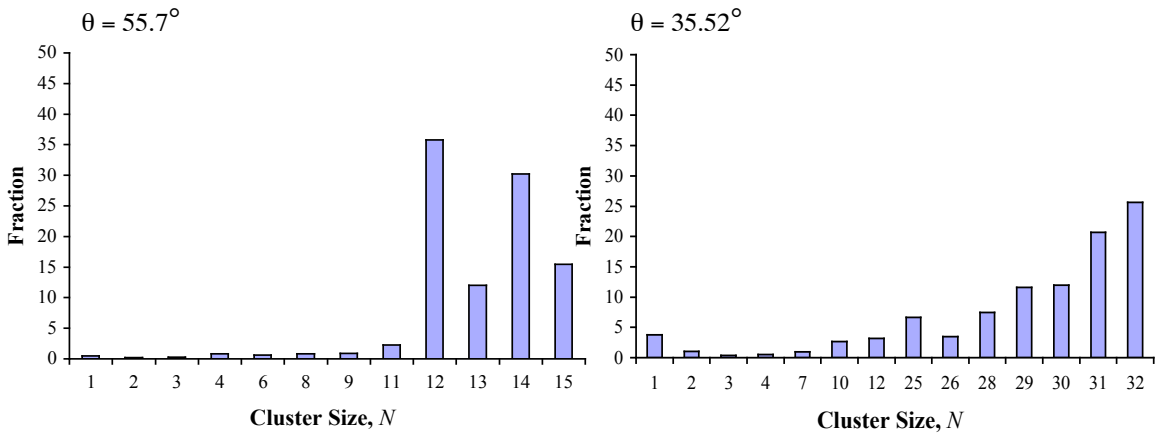

(c)                                           (d)

**FIG.5.**

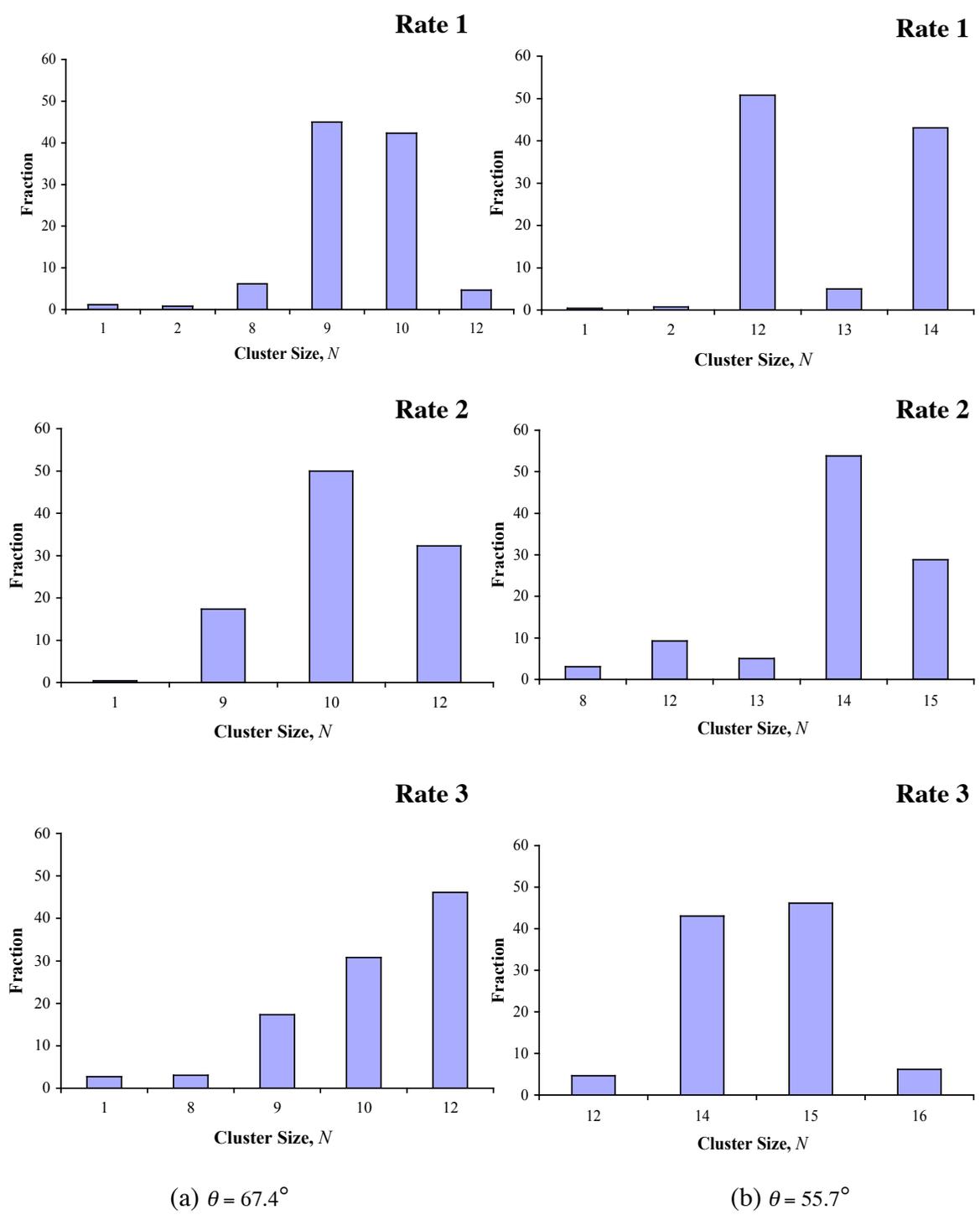

(a) $\theta = 67.4°$

(b) $\theta = 55.7°$

**FIG.6.**

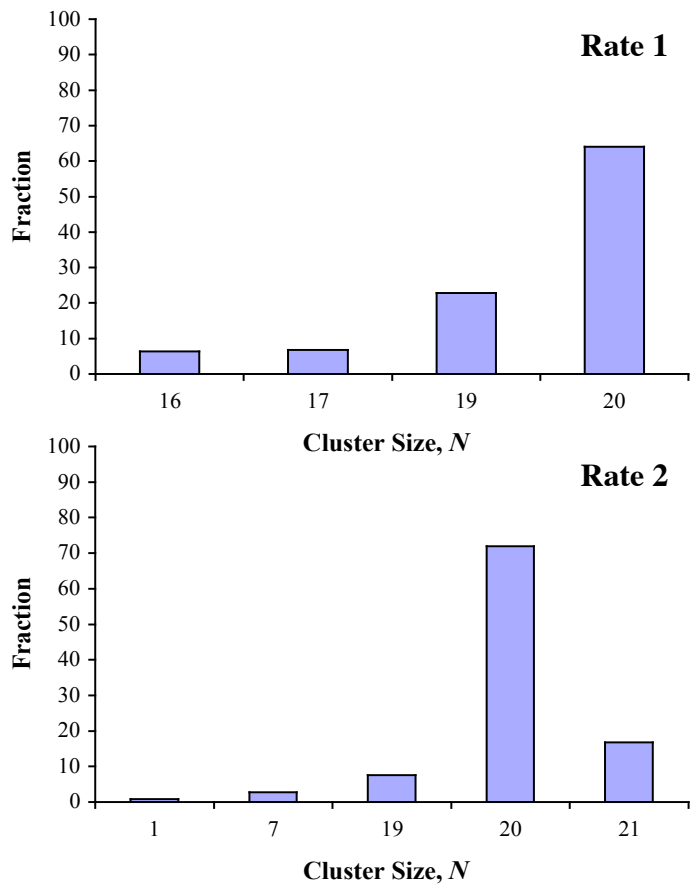

(c) $\theta = 45.3°$

**FIG.6. (continued)**

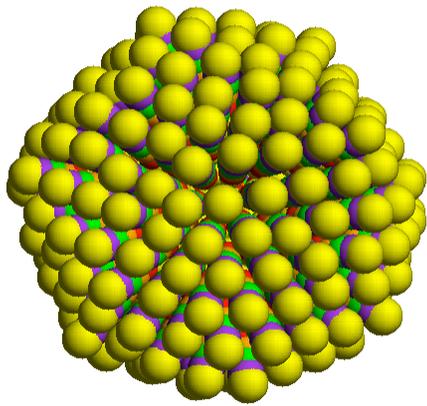
(a)

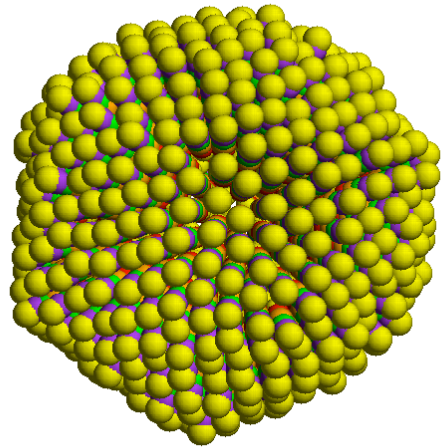
(b)

**FIG.7.**

**TABLE I**

| Size | $N = 6$ | | $N = 9$ | | $N = 12$ | | $N = 13$ | | $N = 14$ | | $N = 20$ | |
|---|---|---|---|---|---|---|---|---|---|---|---|---|
| | $\theta_{min}$ | $\theta_{max}$ | $\theta_{min}$ | $\theta_{max}$ | $\theta_{min}$ | $\theta_{max}$ | $\theta_{min}$ | $\theta_{max}$ | $\theta_{min}$ | $\theta_{max}$ | $\theta_{min}$ | $\theta_{max}$ |
| Rate 0 | 75.2 | 98.9 | 64 | 77.4 | 53.5 | 67.4 | 54.8 | 57.4 | 50.3 | 56.1 | 42.8 | 47.8 |
| Rate 1 | 72.3 | 100.7 | 64 | 78.8 | 54.1 | 68.8 | 53.5 | 58.7 | 50.3 | 57.4 | 42.8 | 48.4 |
| Rate 2 | 76.6 | 98.9 | 66.7 | 77.4 | 54.8 | 68.1 | 56.1 | 56.1 | 51.6 | 57.4 | 43.4 | 48.4 |